\documentclass[preprint2]{aastex}
\begin{document}                  
\title {Two-temperature coronal flow above a thin disk}
\author{ B. F. Liu}
\affil{Yukawa Institute for Theoretical Physics, Kyoto University,
Kyoto 606-8502, Japan}
\affil{Yunnan Observatory, Chinese Academy of Sciences, P.O.Box 110, Kunming 650011, China}
\email{bfliu@yukawa.kyoto-u.ac.jp}
\author{S. Mineshige }
\affil{Yukawa Institute for Theoretical Physics, Kyoto University,
Kyoto 606-8502, Japan}

\author{F. Meyer and E. Meyer-Hofmeister}
\affil{Max-Planck-Institut f\"ur Astrophysik, Karl-Schwarzschild-Str.1,
D-85740 Garching, Germany}

\and 
 
\author{T. Kawaguchi}
\affil{Department of Astronomy, Faculty of Science, Kyoto University,
Kyoto 606-8502, Japan}
 
\begin{abstract}
We extended the disk corona model (Meyer \& Meyer-Hofmeister 1994; 
Meyer, Liu, \& Meyer-Hofmeister 2000a) to the inner region of galactic nuclei 
by  including different temperatures in ions
and electrons as well as  Compton cooling. 
We found that the mass evaporation rate and hence the fraction of accretion
energy released in the corona depend strongly on the rate of incoming mass flow
from outer edge of the disk, a larger rate leading to more Compton cooling,
less efficient
evaporation and a weaker corona. We also found a strong dependence on
the viscosity, higher viscosity leading to an enhanced mass flow in
the corona and therefore more evaporation of gas from the disk below.
If we take accretion rates in units of the Eddington rate
our results
become independent on the mass of the central black hole.   
The model predicts weaker contributions to the hard X-rays for
objects with higher accretion rate like narrow-line Seyfert 1 galaxies
(NLS1s), in agreement with  observations.
For luminous active galactic nuclei (AGN) strong Compton cooling in the innermost corona
is so efficient that a large amount of additional heating is required 
to maintain the corona above the thin disk.
\end{abstract}
\keywords {accretion, accretion disks--galaxies:
nuclei--X-rays:galaxies--X-rays:binaries}

\section{Introduction}
It is commonly thought that AGN are
composed of a central supermassive black hole and a thin equatorial
accretion disk.
Spectral features of AGN, 
the big blue bump, the soft X-ray excess
and hard X-ray tails,
and the 6.4 keV fluorescent iron lines show strong observational
evidence for hot gas in the neighborhood of the inner accretion
disk (e.g., Mushotzky, Done, \& Pounds 1993; Zdziarski 1999). 
Soft disk photons are Compton up-scattered by the optically thin hot plasma
and cause the power-law hard X-ray component in the spectrum.
The relative strength of the hard X-ray component with respect
to the UV--soft X-ray component, leading to the classification as soft
or hard spectrum, is ascribed   
to the dominance of either the thermal emission from the accretion disk 
or of the thermal Comptonization in a hot plasma. 
The iron lines are explained by reflection of the hard X-rays from the
underlying disk (e.g., Fabian et al. 2000; Reeves et al. 2001;
	   Gondoin et al. 2001). 
 
Theoretical simulations successfully reproduce the observed spectra
by assuming a certain fraction of gravitational 
energy released in the hot gas (e.g. Haardt \& Maraschi 1991; 
Kawaguchi, Shimura, \& Mineshige 2001) and/or  a
certain spatial distribution of the hot gas (e.g. Merloni \& Fabian 2001a). 
However, without understanding the
coexistence of disk and corona, the previous models require
additional free parameters  besides 
the mass of the black hole  and the accretion rate for modeling the observed spectra.   

On the other hand, mass evaporation from the underlying cool disk to
the hot corona
provides a mechanism for the change of the accretion disk  in 
the outer region to the coronal flow/ADAF in the inner region. 
Such a model (Meyer \& Meyer-Hofmeister 1994, Meyer et
al. 2000a) describes the physics of the process how mass and energy
exchange and how the hot corona exists above the cool disk. Further
the model allows to explain self-consistently whether and where the
accretion changes from the thin disk to the hot coronal flow/ADAF
in dependence on the given accretion rate. 
The model explains the soft/hard spectral transitions in black
hole X-ray binaries (Meyer, Liu, \& Meyer-Hofmeister 2000b)
and the truncation of the thin disk in soft X-ray
transients and some galactic nuclei (
Liu et al. 1999; Liu \& Meyer-Hofmeister 2001). It provides a scenario
for disk evolution and outburst cycles of soft X-ray
transients (Meyer-Hofmeister \& Meyer 1999). In  these applications
of the evaporation model the inner disk is truncated due to a
relatively low accretion rate.

If we apply the disk corona model to AGN, the inferred
high mass accretion rate in these luminous systems makes it possible for
the disk to extend inward to the last stable orbit since the
high accretion rate tends to prohibit the disk from being depleted. 
Then  the  hard X-ray radiation originates from  re-processing in 
the hot corona.
Such an inner corona has very high temperature and low density,
and hence electrons and ions are no longer well coupled if all the
heating goes to ions and only Coulomb collisions transfer energy from
ions to electrons.  
Photons from the thin disk are Compton up-scattered in the coronal
region close to the central black hole. 
This process influences the coronal structure.
Our former applications of the evaporation model adequate for
equal ion and electron temperature 
(one-temperature coronae) would not be 
a good approximation anymore for the innermost regions of the black hole
accretion system. A two-temperature treatment has to be developed.

Recently a semi-analytical approach based on the same physics is carried
out in detail by
R\'o\.za\'nska \& Czerny (2000). In their investigation the corona is
regarded as a layer with constant pressure,
 which means that the
frictional heating is higher than  that in the real corona
where the pressure
scale height is much smaller as shown by our numerical computations 
(Meyer et al. 2000a).
This approximation and further assumptions in their work lead to quite
a difference in the detailed results.
A recent investigation by Spruit \& Deufel (2001) concerns the heating
of the innermost disk region by hot ions and the resultant evaporation
of matter from the disk.

In the present study we took into account the decoupling of electrons and
ions and the Compton cooling, extending the
previous disk evaporation model (Meyer et al. 2000a)
to small distances from the black hole. 
We calculated the vertical structure of the corona numerically,
attempting to describe the hot
corona self-consistently in equilibrium with the cool disk below.
We also
investigated the dependence on the viscosity value chosen for the hot gas.
We found that the evaporation efficiency is higher for larger viscosity,
that means the mass flow in the corona is higher and consequently the 
mass accretion rate in the underlying disk is lower. Our new results
are of interest with regard to the
observed spectra and  the spectral variations of AGN and galactic
black hole candidates. In Sect.2 we briefly describe the
model. The detailed computational results are presented in Sect.3 . In
Sect.4 we discuss the dominant accretion form at different accretion
rates and the possible coexistence of disk and corona
in the innermost region of AGN. Our conclusions are given in Sect.5.

\section{The model}
We consider a hot corona above a geometrically thin
standard disk around a central black hole. 
Heat released by friction in the corona flows down into lower cooler and
denser corona. There it is radiated away if the density is
sufficiently high. If the density is too low, cool matter is heated up
and joins the coronal gas. We call this process
evaporation.  This goes on until an equilibrium density is established. 
The gas evaporating into the corona still retains angular
momentum, it will differentially rotate around the central object
similar to a standard accretion disk. By friction the gas looses angular
momentum and drifts inward thus continuously drains mass from the corona
towards the central object. This is compensated by a steady mass evaporation
flow from the underlying disk.
The process is driven by the gravitational potential energy released by 
friction in the form of  heat in the corona. 
Therefore, mass is accreted to
the central object partially through  corona (evaporated part) and
partially through disk (the left part of the supplying mass). 
     
\subsection{The equations}
In our earlier study (Meyer et al. 2000a) the equations for the
corona above the disk are derived. The approximations used to describe
the complex physics are discussed there in detail.
For the application to coronae near the black hole we now take
additionally into account  
different ion and electron temperature and hence separate energy
equations for ions and electrons.
Also Compton cooling is considered.

We list the equations as follows,
 
Equation of state
\begin{equation}
P=n_{i}\kappa T_{i}+ n_{e}\kappa T_{e}=
{\Re \rho \over 2\mu} (T_{i}+T_{e}),
\end{equation}
where we assumed a standard chemical composition ($X=0.75, Y=0.25$)
for the corona. The average molecular weight is then 
$\mu=0.62$. For simplicity we take the number density
of ions $n_{i}$ equal to that of electrons $n_{e}$, which is
strictly true only for a pure hydrogen plasma.

Equation of continuity 
\begin{equation}\label{e:cont}
{d\over dz}(\rho v_z)={2\over r}\rho v_r -{2z\over r^2+z^2}\rho v_z.
\end{equation}

$z$-component of momentum equation 
\begin{equation}\label{e:mdot}
\rho v_z {dv_z\over dz}=-{dP\over dz}-\rho {GMz\over (r^2+z^2)^{3/2}}.
\end{equation}

For the derivation of the above two equations see Eqs. (39) and (34) in
Meyer et al. (2000a). The term 
$-\frac{2z}{r^2+z^2}\rho v_z$
is due to the gradual change of the channel ($1+z^2/r^2$ the channel
cross section) of the ascending gas from a 
cylindrical to a spherical expanding shape at large height $z$.

There are now two energy equations (compare Eq. (46) Meyer et
al. (2000a) for the one-temperature case), one for ions and one for
electrons, and the cooling and heating processes in such a hot corona
have to be included.

1) For ions the energy balance is between viscous heating, 
   cooling by collision with electrons and radial and vertical
   advection.

\begin{equation}\label{e:energyi}
\begin{array}{l}
{d\over dz}\left[\rho_{i} v_z\left({v^2\over
2}+{\gamma\over\gamma-1}{P_{i}\over\rho_{i}}
-{GM\over\left(r^2+z^2\right)^{1/2}}\right)\right]\\
={3\over 2}\alpha P\Omega- q_{ie}\\
+{2\over r}\rho_{i} v_r 
\left({v^2\over 2}+{\gamma\over\gamma-1}{P_{i}\over\rho_{i}}-{GM\over\left(r^2+z^2\right)^{1/2}}\right)\\
-{2z\over r^2+z^2}\left[\rho_{i} v_z\left({v^2\over 2}+{\gamma\over\gamma-1}{P_{i}\over\rho_{i}}-{GM\over\left(r^2+z^2\right)^{1/2}}\right)\right],
\end{array}
\end{equation}
where the radial velocity is approximated as $v_r\approx -\alpha
 V_s^2/\Omega r$ ($V_s$ sound speed).
 The friction is assumed to be proportional to
the total pressure, and thus the viscous heating rate per unit volume
is $q_{\rm vis}={3\over 2}\alpha P\Omega$ as appears in the equation.  
Here we assume that all viscous heating goes to ions as usual
(though there is a controversy on possible viscous heating
on electrons (e.g. Bisnovatyi-Kogan \& Lovelace 1997; Quataert \&
 Gruzinov 1999)) and the ion energy  is  transferred to electrons by
 Coulomb coupling alone. Thus, 
the exchange rate of energy between electrons
and ions for non-relativistic approximation is (Stepney 1983),
\begin{equation}\label{e:qie}
\begin{array}{ll}
{q_{ie}}&=\left(2 \over \pi\right)^{1/2}{3\over 2}{m_e\over
m_p}\ln\Lambda \sigma_T cn_{e}n_{i}(\kappa T_i-\kappa
T_e){1+T_*^{1/2}\over T_*^{3/2}}\\
&=3.59\times 10^{-32}n_e n_i(T_i-T_e){1+T_*^{1/2}\over T_*^{3/2}},
\end{array}
\end{equation}
with $$ T_*={\kappa T_e\over m_e c^2}\left(1+{m_e\over m_p}{T_i\over
T_e}\right),$$
and $G$ gravitational constant, $m_p$ and $m_e$ proton and 
electron mass, 
$\sigma_T$ Thomson scattering cross section,
$c$ speed of light, $\kappa$ Boltzmann constant, and Coulomb
logarithm $\ln\Lambda$  taken as 20.

2) For electrons the energy balance is determined by the processes of
 heating by
   collision with ions, cooling by Bremsstrahlung, Compton
   cooling and vertical
   thermal conduction (the radial one is neglected since
  $\partial T_e / \partial r \ll \partial T_e / \partial z$), as well as
   advection.   

\begin{equation}\label{e:energye}
\begin{array}{l}
{d\over dz}\left[\rho_{e} v_z\left({v^2\over
2}+{\gamma\over\gamma-1}{P_{e}\over\rho_{e}}
-{GM\over\left(r^2+z^2\right)^{1/2}}\right)+F_{c}\right]\\
= q_{ie}-n_e n_iL(T_e)-{q_{\rm comp}}\\
+{2\over r}\rho_{e} v_r
\left({v^2\over 2}+{\gamma\over\gamma-1}{P_{e}\over\rho_{e}}
-{GM\over\left(r^2+z^2\right)^{1/2}}\right)\\
-{2z\over r^2+z^2}\left[\rho_{e} v_z\left({v^2\over 2}+{\gamma\over\gamma-1}{P_{e}\over\rho_{e}}
-{GM\over\left(r^2+z^2\right)^{1/2}}\right)+F_{c}\right],
\end{array}
\end{equation}
where $n_e n_i L(T_e)$ is 
the bremsstrahlung cooling rate and ${q_{\rm {comp}}}$  the Compton
cooling rate,
\begin{equation}\label{e:comp}
 {q_{\rm comp}}={4\kappa T_e\over m_e
c^2}n_e\sigma_T c{aT_{\rm eff}^4\over 2},
\end{equation}
with $T_{\rm eff}$ the effective temperature of the main body of the disk,
$$ T_{\rm eff}^4={3G M \dot
M_{d}\over 8\pi r^3 \sigma}, $$ 
here we have added a factor 1/2 to the the energy density of a
black-body photon field in Eq.(\ref{e:comp}) since  
photons from the underlying disks only cover half of the sky of
electrons in the  optically thin corona (compare with Rybicki \& Lightman 1979),
and $F_c$  the thermal conduction (Eq. (49) in Meyer et al. (2000a)),
\begin{equation}\label{e:fc}
F_c=-\kappa_0T_e^{5/2}{dT_e\over dz},
\end{equation}
with $a$
radiative constant, $\sigma$ Stefan-Boltzmann constant, 
$\kappa_0=10^{-6}$g cm  ${\rm s}^{-3} {\rm K}^{-7/2}$ for fully
ionized plasma (Spitzer 1962), 
and $\dot M_{d}$ mass flow rate in the disk.

From Eq.(\ref{e:energyi}) and
Eq.(\ref{e:energye}) we obtain an energy equation 
in the same form as for our one-temperature model,
but with a new term concerning the Compton cooling,
\begin{equation}\label{e:energyt}
\begin{array}{l}
{d\over dz}\left[\rho v_z\left({v^2\over 2}+{\gamma\over\gamma-1}{P\over\rho}
-{GM\over\left(r^2+z^2\right)^{1/2}}\right)+F_c\right]\\
={3\over 2}\alpha P\Omega-n_e n_iL(T_e)-{q_{\rm comp}}\\
+{2\over r}\rho v_r
\left({v^2\over 2}+{\gamma\over\gamma-1}{P\over\rho}
-{GM\over\left(r^2+z^2\right)^{1/2}}\right)\\
-{2z\over r^2+z^2}\left[\rho v_z\left({v^2\over 2}+{\gamma\over\gamma-1}{P\over\rho}
-{GM\over\left(r^2+z^2\right)^{1/2}}\right)+F_c\right]
\end{array}
\end{equation}
For convenience the two energy equation 
Eq.(\ref{e:energyi}) and Eq.(\ref{e:energyt}) 
are used in the numerical calculations instead of Eq.(\ref{e:energyi})
and Eq.(\ref{e:energye}).

We want to point out that the one-zone approximation is taken for
deriving the above equations. For a given distance $r$ we choose a region  
between  $r_1$ and $r_2$ so that
\begin{equation}
\pi (r_2^2-r_1^2)=\pi r^2
\end{equation}
and 
\begin{equation}
r={\int_{r_1} ^{r_2} r^\prime \cdot 2\pi r^\prime dr^\prime \over \pi (r_2^2-r_1^2)},
\end{equation}
i.e. $r_1=0.72r$ and $r_2=1.23r$.
In the one-zone region the radial advection is
     taken related to the difference between mass inflow from the
outer neighboring
     regions and outflow towards the inner neighboring regions,
\begin{equation}
{(2\pi r \Delta z \rho v_r)_{r_2}-( 2 \pi r \Delta z \rho v_r)_{r_1} \over
2\pi r \Delta z (r_2-r_1)} \approx -{\rho v_r \over (r_2-r_1)} 
\approx -{2 \over r}\rho v_r 
\end{equation}
This approximation
for the net radial flow requires the mass outflow larger than the inflow.
In the following sections we will see that the evaporation rate
reaches a maximum and  doesn't increase any more 
(see Fig.~\ref{f:mdot-r}) inside the maximal evaporation region. This implies our
results are inconsistent in the inner region
and we need to reconsider the 
radial flow. For this we study the radial
flow more accurately in Sect.\ref{s:discussion1}. It turns out that the effect of the radial advective
     loss is very small to the inner structures, the
corresponding terms  in the continuity  and energy equations are not
very important. Therefore, our results based on one-zone approximation 
are still consistent.

\subsection{Boundary conditions}
The 5 differential equations, Eqs.(\ref{e:cont}),
(\ref{e:mdot}),(\ref{e:energyi}),
(\ref{e:fc}) and (\ref{e:energyt}), contain 5 variables $P(z)$, $T_i(z)$,
$T_e(z)$, $\dot m (z)(\equiv \rho v_z)$ and  can be solved with 5 boundary
conditions.

At the lower boundary $z_0$  (the interface of disk and corona),
there is no heat flux and the temperature
of the gas should be the effective temperature of the accretion
disk.  Former investigations (Liu, Meyer, \& Meyer-Hofmeister 1995) show
that the coronal temperature increases from effective temperature to
$10^{6.5}$K in a very thin layer and thus the lower boundary
  conditions can be reasonably approximated (Meyer et al. 2000a) as,
\begin{equation}
T_e=10^{6.5}K,\  T_i=T_e,\  {\rm and} 
\  F_c=-2.73\times 10^6 P\
{\rm at}\  z=z_0
\end{equation}
At infinity, there is no artificial confinement and hence no pressure.
This requires sound transition at some height $z=z_1$, which is taken
as our upper boundary. There is no heat flux from/to infinity
either. We then constrain the upper boundary as,
\begin{equation}
F_c=0\  {\rm and}
 \  v_z^2=V_s^2\equiv P/\rho={\Re\over 2\mu}(T_i+T_e)\  {\rm at}\  z=z_1 
\end{equation}
With such boundary conditions, we have to assume lower boundary values
for  $P$ and $\dot m$ to begin the
integration along $z$, if the trial $P$ and $\dot m$ fulfill the upper boundary
conditions, we find the solutions.  
\section{Computational results}

We solved the differential Eqs.(\ref{e:cont}),
(\ref{e:mdot}),(\ref{e:energyi}),
(\ref{e:fc}) and (\ref{e:energyt}) using the Runge-Kutta
method. The technical details of the integration procedure are similar
to those in our previous study (Meyer et
al. 2000a). In the present investigation standard parameters were taken 
for the black hole mass $M=10^8M_\odot$ and the viscosity coefficient
$\alpha$=0.3. 
The Compton cooling by photons depends on the
radiation from the underlying disk and hence on the mass flow rate
in the disk. This is an additional parameter on which the evaporation
efficiency depends and  this parameter is different in different
astrophysical objects. In order to study the  coronal 
properties without involving the mass accretion rate in the disk, 
 we first investigate the influence of viscosity without 
Comptonization, that means we assumed a very low mass accretion rate in
the disk. The results are presented in Sects.~\ref{s:vert}---\ref{s:visc}. 
The computational results for the effect of Compton scattering
are discussed in Sect.~\ref{s:comp}, the relevance for
corona around stellar-mass black holes in Sect.~\ref{s:stel}. 

\subsection{Vertical structure of the corona}\label{s:vert}
\begin{figure}

\plottwo{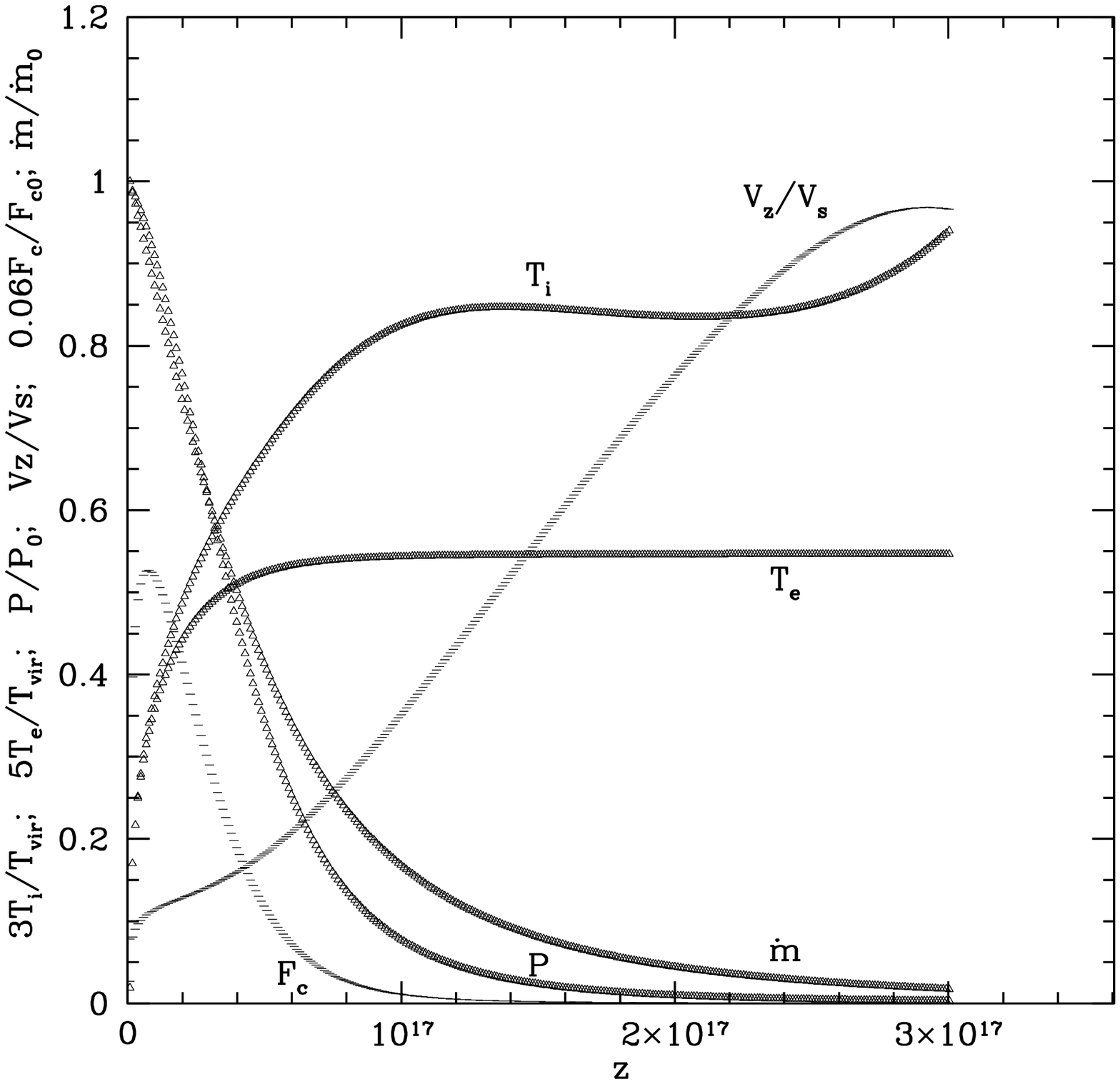}{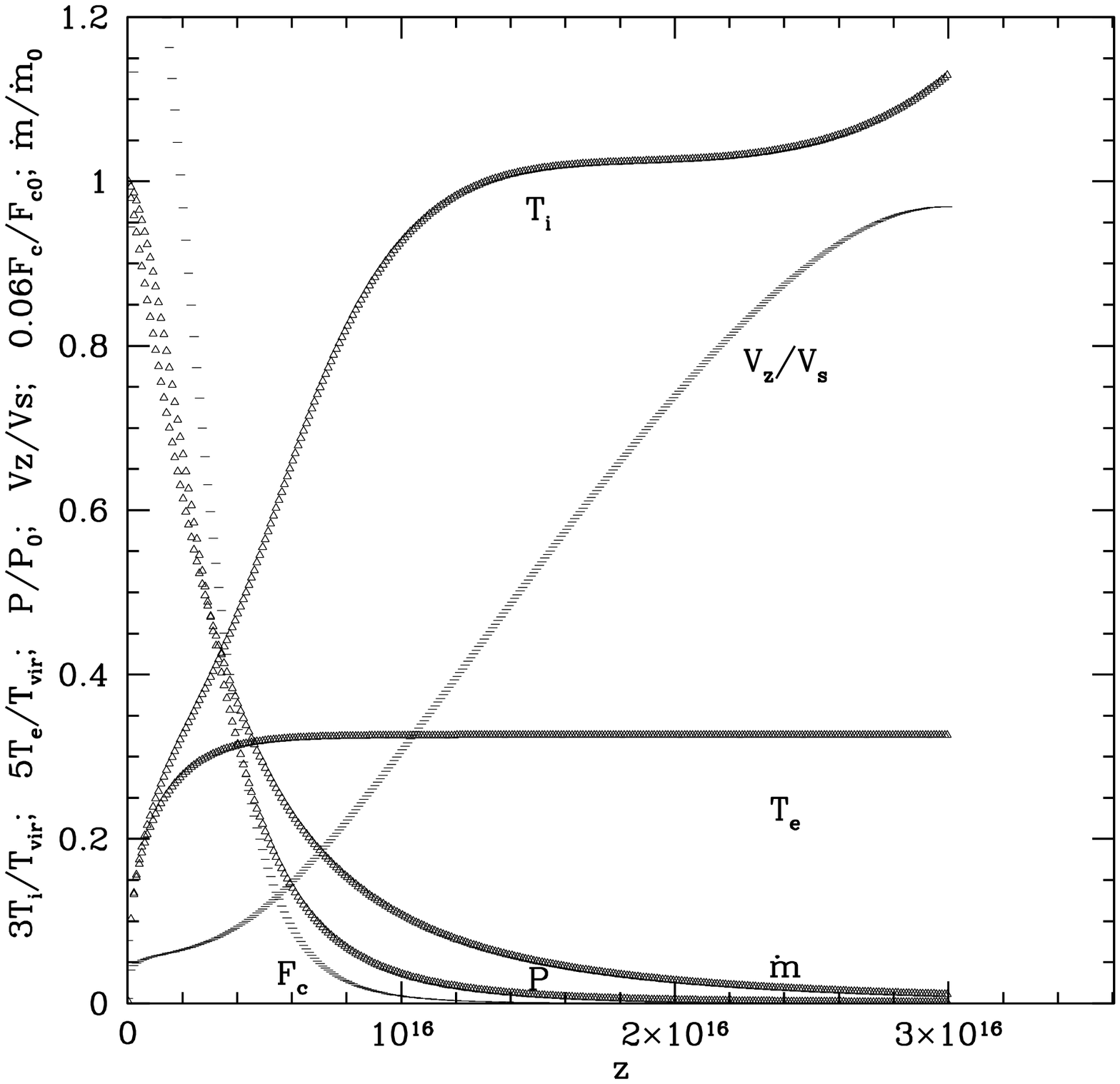}
\caption[]{\label{f:vert} Vertical structure of the corona 
at different distances from a central black hole of
mass  $10^8M_\odot$,  at $r=3400 r_g$(left panel) and 
$r=340 r_g$ (right panel). 
Distributions of the following quantities: Temperatures of ions $T_i$ 
and electrons $T_e$ in units of 0.2$T_{\rm {vir}}$,
($T_{\rm {vir}}$ virial temperature); 
pressure $P$, 
thermal conductive flux $F_c$ and vertical mass flow $\dot m=\rho v_z$
scaled in their values at the
lower boundary.
Steep variations of $P$, $T_i$, $T_e$, and $\dot m$ occur in the lower
layers with $z<r$, above these transition layers the corona is
quasi-constant.}
\end{figure}

Fig.~\ref{f:vert} shows the vertical structure of the corona at
distances $r=3400r_g$ and $r=340r_g\ (10^{16}$cm for $M=10^8M_\odot$), where 
$r_g=2.95\times 10^{13}(M/10^8M_\odot$)cm is the Schwarzschild
radius, 
and $T_{\rm vir}=GMm_p\mu/\kappa r=3.37\times 10^{12}(r/r_g)^{-1}$K
the virial temperature.
The variation with height $z$ of the  variables,   $P(z)$, $\dot m(z)$,
$F_c(z)$,
$T_e(z)$,  $T_i(z)$,  and $V_z/V_s$, is similar to that in our
previous computations.
The temperature, pressure, density keep almost invariable
in the corona form $z\sim r$ to the upper boundary ($z\la 3r$).
Steep changes occur in the lower corona  ($z\la r$), where the temperature
increases from a chromospheric value ($T_e = T_i \approx 10^{6.5}$K)
to a coronal value ($T_i \sim (0.2-0.3)T_{\rm vir}$) and the pressure
decreases rapidly.  The density, $\rho\propto P/T$,
drops even more steeply than temperature in the lower corona.

At large distance $r\sim 3400r_g$
electrons in the corona are well coupled with ions, 
and thus electron and ion temperatures are nearly the same,
$T_e\approx 10^8$K. At a distance close to the black hole, Coulomb
coupling between ions and electrons becomes 
weak due to low density and high temperature, and the temperature in
ions (e.g. $T_i\approx 10^{11}$K at $r=10r_g$)
is about 100 time higher
than that in electrons.

\subsection{Distribution of the temperature versus distance}
\begin{figure}
\plotone{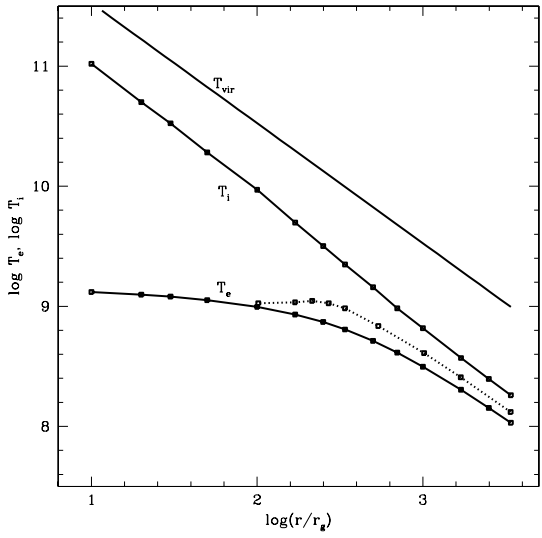}
\caption[]{\label{f:T-r} Temperature distribution versus
 distance from the black hole. Solid lines lower to upper: 
distributions of  electron and ion temperature $T_e$ and $T_i$,
and virial temperature  
$T_{\rm vir}$, respectively. Dotted line: temperature distribution
of a 1-T model (Liu \& Meyer-Hofmeister 2001). 
The ion temperature increases towards the central black hole, the electron
 temperature stays 
around $10^9$K in the inner region. The distribution resulting from
the 2-T  disk corona model is almost the same as that from 
a typical ADAF.}
\end{figure}

Fig.~\ref{f:T-r} shows the change of the coronal temperatures with
distance.
The ion temperature increases towards the central black hole, 
from values of about $\frac{1}{5}T_{\rm vir}$ 
in the outer region (e.g. $T_i\approx 1.8\times 10^8$K at $r=3400r_g$)  
to $\frac{1}{3}T_{\rm vir}$ close to the black
hole (e.g. $T_i\approx 10^{11}$K at $r=10r_g$), 
while the electron temperature hardly increases beyond the value of $T_e\sim
10^9$K from the distance $r\sim 100r_g$ inward to the black hole.  
The temperature distribution of ions differs from that of electrons 
since the viscous heat of
ions can hardly be transferred by collisions to electrons and is
only  transferred to the entropy of ions. The fraction of
heat turned into entropy 
 becomes larger at small distance.
 The electrons, however, can efficiently
 cool by thermal conduction and radiation 
and hence keep their temperatures almost constant at decreasing
distance in the inner region.
Such distributions of ion and electron temperature versus distance in
the corona are almost the same as those in a 
typical ADAF (Narayan \& Yi 1995;  Nakamura et al. 1997; 
Manmoto, Mineshige, \& Kusunose
1997). The evaporation  of mass is inefficient in the
inner accretion region and the viscous heat is transferred to
entropy. The corona is essentially identical with an ADAF.

For comparison the temperature distribution resulting from our previous
one-temperature (1-T) model is also plotted in Fig.~\ref{f:T-r} (dotted
line). From the outer regions  to the distance of most
efficient evaporation ($r\sim 340 r_g$) 
the  1-T temperature lies between the temperatures of ions and
electrons of two-temperature (2-T) model, actually the three
temperatures are not much different.
In the innermost region the one-temperature solutions are not
adequate anymore unless there exists much more efficient coupling than
Coulomb coupling. 

\subsection{Distribution of evaporation rate versus distance}

\begin{figure}
\plotone{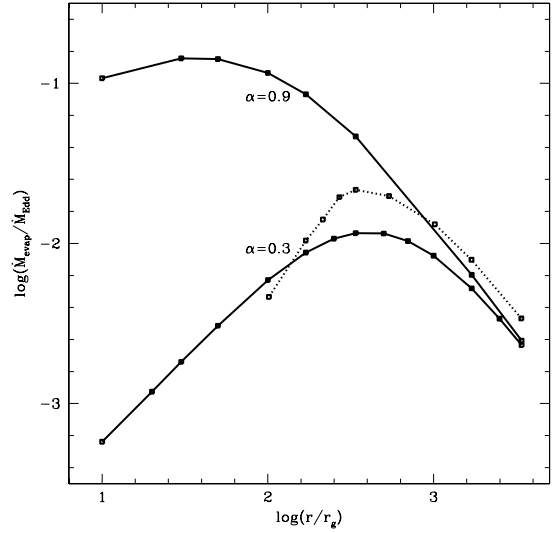}
\caption[]{\label{f:mdot-r}Evaporation   rate versus
distance.
Solid lines: results from the 2-T model for a viscous coefficient $\alpha=0.9$
(upper line) and $\alpha=0.3$ (lower line); dotted line: result from
the earlier computations using a 1-T model, $\alpha=0.3$.
The mass evaporation rate increases more than 10 times when the viscosity
increases 3 times. The evaporation rate
resulting from 2-T model is a little smaller than that from the
1-T model since a small part
of heat remains in ions to maintain a slightly higher temperature
and less heat goes to electrons (lower electron temperature)
and is conducted downwards for evaporation.}
\end{figure}

We define the mass evaporation rate from our one-zone region as 
$\dot M_{\rm evap}(r)\equiv 2\pi r^2\dot m_0(r),$
where $\dot m_0(r)$ is the evaporation rate from unit surface area 
at the interface of the corona and the chromosphere (i.e. the lower boundary).
Fig.~\ref{f:mdot-r} shows the rate of mass evaporating 
from the disk to the
corona versus distance.
The evaporation rate $\dot M_{\rm evap}(r)$ increases with
decreasing distance $r$, reaches a maximum (hump) of $\dot M_{\rm
hp}=0.012 \dot M_{\rm Edd}$ ($\dot M_{\rm Edd}=10L_{\rm Edd}/c^2$ 
Eddington rate)
at $r_{\rm hp}\sim 340 r_g$, and
decreases towards the black hole. The occurrence of 
evaporation hump is an important characteristic of our disk corona
model.  The maximum of the coronal flow rate
is caused by the change in the 
physical processes that remove the frictional heat. At large distances the
coronal heat is balanced by the inward advection and wind loss.
Downward heat conduction and subsequent radiation in 
the denser lower region play a minor role. Thus, the evaporation is weak.
At small distances thermal conduction removes an
increasing part of the viscous heat and becomes dominant with rising
temperature.  This downward conducted viscous heat then evaporates the
cool gas at the chromosphere  and leads to efficient
radiation. Therefore, the mass evaporation rate in the corona
 increases continuously  with decreasing radius until the radius
$r_{\rm hp}$ 
is reached. At distances $r<r_{\rm hp}$, the electrons are no longer well
coupled with ions in the corona and downward thermal conduction is
less important; on the other hand, since the density increases towards
the black hole, the heat at the lower coronal layer
can easily be released by radiation in the innermost region, 
and thus evaporation is not active.
 
Also shown 
in Fig.~\ref{f:mdot-r} (dotted line) is the distribution of the
evaporation rate $\dot M_{\rm evap}(r)$ 
 from our previous 1-T model ($\alpha=0.3$). The maximum lies at the
same distance from the black hole, but
its value is one time
larger than that of the 2-T model. The 
difference of evaporation efficiency between the 2-T and the 1-T model 
originates from the temperature difference,
the slightly lower electron  temperature in the 2-T model leads to
less thermal conduction and resultant evaporation.   
 A small part of heat remains in ions to keep the ion temperature
slightly  higher than  the temperature in the 1-T model instead of going to
 electrons and evaporating disk gas.

The gas evaporated from the disk into the corona flows inward through
the corona towards the black hole, except for a small fraction leaving by
wind.
Therefore the coronal mass flow rate at each distance $r$
is the amount of gas evaporated from the disk
at distances further outward up to $r$. 
Since the mass evaporation rate has a steep hump, the coronal
flow at radius $r$ outside the hump is approximately $\dot M_{\rm
evap}(r)$,
 while at radius inside the hump it is a little bit higher than the maximal evaporation
rate $\dot M_{\rm hp}$.  

\subsection{Energy flows in the corona}

\begin{figure}
\plotone{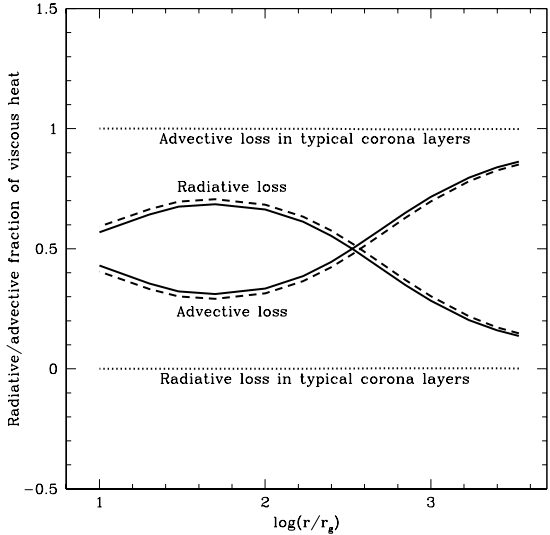}
\caption[]{\label{f:energy-r} 
Variation of the energy losses with distance $r$ from the black hole.
Fraction of viscous heating going into radiation and fraction removed
by advection. Solid lines: rates integrated over the total corona from
the lower boundary ($z_0$, chromosphere) to the upper boundary (sound
transition). Dashed lines: rates integrated over the transition layer
only, from $z_0$ to $z=r$. Dotted lines: rates from $z=r$ to the
upper boundary. The comparison of solid and dotted lines show that
the processes in the lower coronal layers are responsible for the higher
radiative loss fraction in inner region and the lower advective
loss fraction in the outer region. Considering the upper coronal
layers only, the viscous heat almost flows inwards as entropy with
little radiation loss (a few thousandths in outer region), which is
quite similar to the ADAFs.} 
\end{figure}

 For the local energy balance 
we consider the processes of viscous heating, radiative cooling, 
radial advection and vertical advection (energy brought into the
corona by gas evaporated from the disk and removed in the wind). 
Considering the coronal structure at a given distance $r$ from the
black hole we integrate the energy loss by radiative cooling and the
loss by vertical and sidewise advective outflow over $z$, the height
of the corona. It is an interesting feature how these losses,
normalized by the release of frictional energy, vary with distance $r$. 
We plotted the values in Fig.~\ref{f:energy-r}. It shows that 
the fraction of viscous heat which is lost by radiation is larger at
small distances than at large distances, while
advection is dominant in outer the region ($r\sim 3400r_g$). It takes away
the major fraction of viscous heat there. At the distance
$r\sim 340r_g$, where evaporation rate reaches its maximum, 
advection and radiation both remove half of the viscous heat
respectively. In the inner region the density in the layers 
at the bottom of the corona is large, thus the  radiation becomes
efficient and takes away a larger fraction of viscous heat than advection.  
This feature seems to be in contrast to the typical ADAFs, in which
the flows are radiation dominated further out and become more and
more advection dominated with decreasing
distance until finally the advective cooling balances the viscous
heating in the innermost region (Nakamura et al. 1997; Manmoto et al. 1997). 
But with a
corona above the thin disk evaporation leads to mass exchange and
therefore the flow is not a constant radial mass flow  
in the corona as in an ADAF.

The detailed calculations of the vertical structure allow to study
the difference in the energy flow in the upper corona as compared to
that in the lower layers (compare also Fig. 4 in Meyer
et al. 2000a).
Radiation and thermal conduction dominate the energy
processes in the very low layer, the heat conducted downwards
by electrons and subsequently radiated away in the lower layers is
provided by the viscous heat produced in  the whole region
from the chromosphere to a height of $z\sim r$. In the upper layers
the pressure is very low and the small amount of viscous heat produced
there is almost completely advected.  Thus, the total radiation
(vertically integrated over the  whole corona) 
predominantly originates from the lower coronal layers. 
Considering the layers from height $z=r$
to the upper boundary, the viscous heat is advected, as shown
in Fig.~\ref{f:energy-r} (dotted lines). The very small fraction of radiative
loss (e.g. a few thousandths at $r\sim 3400r_g$), decreases
towards the black hole. In this sense, the corona is
advection-dominated,
consistent with the ADAF except that our corona model  
also includes vertical advection flow.

Comparing the energy flow in our computed corona 
with an ADAF, we find that the  difference 
between the disk corona and the ADAF is in the
supply of accretion gas. In the ADAF, accreting matter streams inward
from the further outward adjacent ADAF region, where the temperature
is also high. While for the coronal flow above the disk,
at $r>r_{\rm hp}$,
the accreting mass is mainly supplied from the underlying disk with a
temperature far lower than that in the corona and the evaporating gas
is heated to a high temperature before it accretes inwards. This needs to
increase the entropy and the potential of the gas by 
``absorbing'' viscous heat, and meanwhile the heat is partially radiated
in the dense lower corona. If we consider the whole vertical region,
 the lower and upper coronal layers, 
the average temperature should be lower than that
of an ADAF with the same mass supply rate, 
and the average radiative cooling is more efficient due to the
structure of the lower coronal layers. After the essential increase of
temperature and corresponding changes of the other quantities in the
lower corona the quantities are nearly constant in the upper layers,
where the gas density is very low and the temperature is rather high.
Such a corona is radiation-inefficient, almost all of the viscous heat
is advected with the accretion flows (dotted lines in Fig.~\ref{f:energy-r}). 
These typical coronal layers  are 
actually like an ADAF, which can also be seen from the temperature
distribution shown in Fig.~\ref{f:T-r}.

\subsection{Effect of viscosity}\label{s:visc}
The dependence of the evaporation rate on the viscosity
is shown in Fig.~\ref{f:mdot-r}. We have chosen a relatively large
viscous coefficient, $\alpha=0.9$, for comparing with the results
for our standard value $\alpha=0.3$. As already predicted by an earlier
analytical study of  this dependence, the influence on the evaporation
efficiency is strong.
For a larger viscosity in the corona the evaporation rate is higher.
The radial location of the maximal rate is shifted inwards for higher
viscosity values. The maximal rate for 
$\alpha=0.9$ is 0.14$\dot M_{\rm Edd}$, about 10 times larger than that for
$\alpha=0.3$, and the maximum occurs at a distance about 30 Schwarzschild
radius, about 10 times smaller than that for $\alpha=0.3$.
The reason is that the larger viscosity leads to more efficient heating 
and an enhanced mass flow  in the corona.
The increased viscous heat is then conducted  downward  by electrons
and has to be radiated away by more efficient evaporation of the gas
in the disk atmosphere. 
On the other hand, faster accretion in the corona tends to decrease
faster the
number density of the corona  and requires more mass supply by
evaporation to keep the energy balance at the interface of corona and
disk. Therefore, the increase in viscosity largely increases mass
evaporation. Similar results were obtained in an investigation using
the 1-T model (Meyer-Hofmeister \& Meyer 2001), where small viscosity
is assumed.

As an important consequence of this dependence on viscosity, the
distance where the mode of accretion via the thin disk 
changes to an ADAF is also affected. For a given mass inflow from the
outer part the disk can extend farther inward for smaller values of
viscosity. Therefore, the emerging spectra should be different
if two systems have a similar mass accretion rate but different viscosity    
in the hot coronal gas (for details see Meyer-Hofmeister \& Meyer
2001). But it is difficult to draw detailed conclusions from the observations.

 \subsection{Effect of Compton cooling}\label{s:comp}
At small distance from the black hole Compton cooling affects the
structure of the corona. The Compton cooling term is included in 
the energy equation (\ref{e:energyt}). Considering the processes
which enter in the energy balance, Compton cooling is the only process
which causes a dependence of the coronal structure on the accretion
luminosity and therefore the mass flow in the underlying disk. 
Otherwise the equilibrium between corona and disk relies only on the
existence of cooler gas below. The dependence on the mass flow in
the disk adds one more parameter to the evaluation of coronal
structure and the evaporation rate. To eliminate this problem 
we  derived the results
assuming a negligible Compton cooling (corresponding to very low mass
flow rate in the disk) in the foregoing sections . 
 
\begin{figure}
\plotone{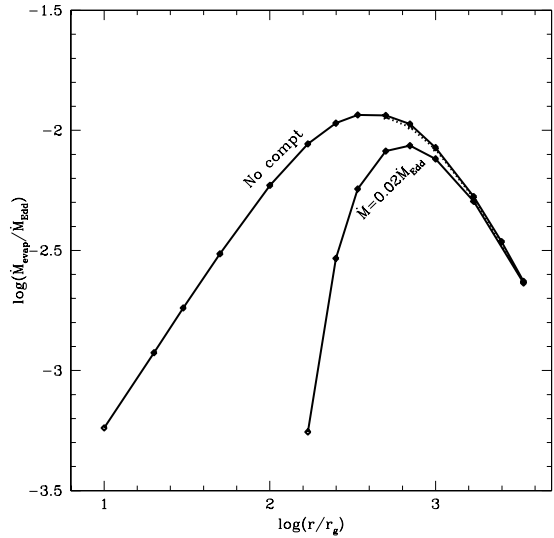}
\caption[]{\label{f:mdotc-r} Effect of Compton cooling on the
evaporation. 
Upper solid line: Evaporation rate without Comptonization (corresponds to a
very low mass flow rate in the thin disk); lower solid line and dotted
line: including
Compton up-scattering of soft photons from the disk with
  a total accretion  rate of $\dot M_{\rm{in}}=0.02 \dot M_{\rm  Edd}$ 
 and $\dot M_{\rm{in}}=0.01\dot M_{\rm  Edd}$, respectively. 
The gas is partially evaporated into a coronal
flow and partially accretes through the disk. The figure shows that
the evaporation can be strongly restrained by Compton cooling, depending on
the mass flow rate in the disk, in particular in the inner region.}
\end{figure}

To investigate the effect of Compton cooling in a disk around a
supermassive black hole with a mass of $10^8M_\odot$ we assumed a rate
of $\dot M_{\rm in}=0.05 M_\odot{\rm  yr}^{-1}$ ($\approx 0.02\dot M_{\rm Edd}$),
corresponding to a bolometric luminosity
$L=2.8\times 10^{44}$ergs${\rm\ s}^{-1}$ (assuming an energy-conversion
coefficient of $0.1$).
 A part of the mass flow goes to the corona through evaporation.
So at a given distance the net mass flow rate in the disk should be 
\begin{equation}
\dot M_{\rm d}(r)=\dot M_{\rm in}-\dot M_{\rm c}(r),
\end{equation}
where $\dot M_{\rm
c}(r)$ is the integrated evaporation rate from outer edge to the 
distance $r$. This relation allows to
evaluate the energy density of soft photons from the underlying disk to be 
scattered in the corona by assuming blackbody radiation. 
We calculated the coronal structure and the evaporation rate for
different distances from large distance towards the central black hole. 
In Fig.~\ref{f:mdotc-r} these results for the assumed total mass flow
rate $0.02\dot M_{\rm Edd}$ and $0.01\dot M_{\rm Edd}$ are shown  
together with the results without taking the
Compton cooling into account (corresponding to a very low mass flow
rate).
As expected  the Comptonization has a strong influence on the
structure of the corona if the mass flow rate in the disk is high.
Since the mass flow rate at distance $r$ in the disk depends on the
evaporation in the further outward regions,  whether the
Compton cooling affects the evaporation rate at $r$ depends on the
accretion in the outward disk plus corona.
At large distances, e.g. $r\sim 10^3r_g$, the corona is hardly
affected by Compton scattering, while in the inner region, Compton
cooling is so strong that it 
overwhelms Bremsstrahlung and removes viscous heat efficiently, 
there is no much need for vertically advective cooling and hence little gas
evaporates from disk to corona.
For the example with mass accretion rate $0.02\dot M_{\rm Edd}$ shown 
in this figure,
the evaporation rate at $r=340r_g$ is  half of the value
without Comptonization, and drops very steeply inwards, reaching a
value of $\sim 15$ times less than the non-Compton one. 
In other words, strong Comptonization in the inner region restrains the
evaporation. The corona then becomes rather weak.

An interesting feature caused by Compton cooling
is the possible shift of the location of the maximal evaporation rate
to a larger distance from the black hole.
Computations for a series of  mass flow rates show that 
from a larger total mass flow rate $0.02\dot M_{\rm Edd}$ to those for
negligible mass flow in the disk, the evaporation rate at given $r$
continuously increases 
and location of the maximal evaporation moves inwards,
as shown in Fig.~\ref{f:mdotc-r}.
For a total mass flow rate from the outer regions of
$\dot M_{\rm in}=5.0\times 10^{-4}\dot M_{\rm Edd}$,
 the Compton cooling does not affect
the evaporation.
For a rate 
$\dot M_{\rm in}=0.01\dot M_{\rm Edd}$, there is only very little 
influence by Compton cooling (see the dotted line in Fig.~\ref{f:mdotc-r}), 
at the distance $r=340r_g$ where the evaporation rate is maximal,
there is already no enough mass flow to supply the efficient evaporation.
For a rate $0.02\dot M_{\rm Edd}$ the changes are correspondingly larger,
and the evaporation rate at $r=340r_g$ is small and already inside
the location of the maximal evaporation rate which moved outward to a
larger distance. 
In our previous papers (Liu et al. 1999, Meyer-Hofmeister \& Meyer
1999) we suggested that the transition from the thin disk accretion
to a coronal flow/ADAF is caused by the mass evaporation. The new
results including Compton cooling by the disk radiation which show a
decrease of the evaporation efficiency in the inner disk regions
imply the transition in the accretion mode for a given mass flow rate
in the disk (a rate below the maximal evaporation efficiency) at a
smaller distance from the black hole. For higher rates
the change between the hard and soft spectral states is expected,
which occurs at 
a lower mass inflow rate than that without Compton cooling.

The effect of Comptonization obviously depends on the total mass accretion
rate. Since large accretion flow 
in disk provides strong seed photons to be Compton scattered in the
corona, the conductive heat at the lower layers of the corona can be
appropriately released by Compton radiation, the evaporation will
decrease
or stop from some distance inwards. 
At very large accretion rates, our computations show that
the Compton cooling is so efficient
that  either  sound transition at the upper boundary is inaccessible, 
or the corona needs 
artificial heat from the upper boundary, in other words, no
evaporation solutions are found. 
This indicates that probably part of the 
coronal gas  falls back to the disk
by efficient Compton cooling in the innermost region.

\subsection{Corona around stellar-mass black hole}\label{s:stel}
 
Since the physics of a disk corona around a stellar-mass black hole
is same as that around a galactic black hole, we performed with the
same code numerical calculations for accretion onto a stellar-mass
black hole of $M=6M_\odot$ to study the influence of evaporation
for adequate mass flow rates. 
We obtained similar results for both vertical and radial structure.
If the distance is measured in Schwarzschild radius $r_g$
and the mass accretion rate in Eddington rate, 
the radial distribution of 
the evaporation rate  and the location of the maximum rate are the same as
those for a supermassive black hole.  The values of temperature at any
distance (in Schwarzschild radius) for both
systems are also the same. As for the vertical structure,
the quantities are  mass independent as long as we scale
the size (distance $r$ and height $z$) in Schwarzschild radius and 
multiply the quantities $F_c$, $P$, $\rho$, $\dot m$ by the mass of the black
hole. This is essential since our equations and boundary conditions
are invariant against scale transformations that leave temperature and
velocities constant. Therefore, the coronal structures are scale free
with respect to the mass of central black holes.

Dimensional analysis of the Compton cooling term and its related
equations show that the coronal structures are still 
independent on the mass of
the black hole even if the Compton cooling is included. This is easily
understood by
comparing the Compton cooling term and viscous heating term in the
energy equation. If we scale
the distance  in Schwarzschild radius 
and the disk accretion rate in Eddington rate we obtain the
dependences of $q_{\rm vis}$ and $q_{\rm comp}$
on the black hole mass as,  $q_{\rm vis}\propto P\Omega\propto PM^{-1}$,
$q_{\rm comp}\propto {T_e\over T_e+T_i}PM^{-1}$. Obviously the
Compton cooling term has the same dependence on $M$ as the viscous
term, implying that the results are also mass independent when
Compton cooling is included.
If we take the rate $0.01 \dot M_{\rm Edd}$ ($0.025M_\odot {\rm yr}^{-1}$ for a $10^8 M_\odot$ black hole)
as upper limit for which the
corona  
is not affected by Compton cooling (the true value should
be slightly smaller  as shown in
Fig.~\ref{f:mdotc-r}),
we  derive an accretion rate of
$2.5 \times 10^{-10}M_\odot {\rm yr}^{-1}$ (or $9.5\times 10^{16}$g$\ {\rm s}^{-1}$)
for a $6M_\odot$ black hole for the
limit. This rate is a little less than that expected for X-ray
transients at spectral transition. This means the hard/soft spectral
transition of X-ray transients is hardly
affected by Compton cooling. The luminosity for which spectral
transitions occur would be only slightly lower with Compton 
cooling included.

Despite the similarity between the corona above a stellar mass and
that above a supermassive black hole (quantities scaled as described)
the spectra would be different even if the mass
accretion rates in unit of Eddington rate are the same. The reason are
the different temperatures in the cool disk around the black holes.
The situation is similar to that of the inner ADAFs with outer thin disks.

\section{Discussion}

\subsection{Validity of the approximation of radial advective loss}\label{s:discussion1}
In Sect.2.1 we already pointed out the radial advection problem. 
For our computation of the 
coronal structure in a given zone of radial extent $r_2-r_1\approx 0.5r$,
we assumed that the mass entering from outer adjacent region is less
than the mass evaporating into this zone, and thus we approximated
the radial mass loss ${1\over r}{\partial r\rho v_r\over \partial
r}\approx -{2\over r}\rho v_r$. Similar approximation was taken 
for the energy advection. Computations based on this one-zone model
show that evaporation does not increase any more in the inner region.
This means that our approximation 
is justified only for distances from the black hole larger
than $r_{\rm hp}$. At smaller distances 
the mass flow in the corona should also include the mass entering from outer
adjacent region besides the evaporation flow, and 
the radial advective loss has to be taken
consistently with the mass flow change in the corona.

At a coronal region inside the hump, an inflow of $\sim
\dot M_{\rm evap}(r_{\rm hp})$ superimposes on
the evaporation flow, increases the coronal density and pressure.  As a
result, the gradient of radial mass flux $-{2\over
r}\rho v_r$, which is contributed by the evaporation mass in the
one-zone region, should be formally
 decreased by a factor of $\eta (r)$ to compensate the increase of the
density, i.e., 
\begin{equation}
{1\over r}{\partial r\rho v_r\over \partial r}\approx -{2\over r} \eta(r)\rho v_r
\end{equation}
with
\begin{equation}
\eta(r)\approx {\dot M_{\rm evap}(r)\over \dot M_{\rm c}(r_2)+\dot M_{\rm evap}(r)},
\end{equation}
where $\dot M_{\rm c}(r_2)$ is the mass flow coming from the outer edge  of
one-zone region and $\dot M_{\rm evap}(r)$ the mass flow evaporating into the 
one-zone region.
The net energy flow is also multiplied by $\eta(r)$.

\begin{figure}
\plotone{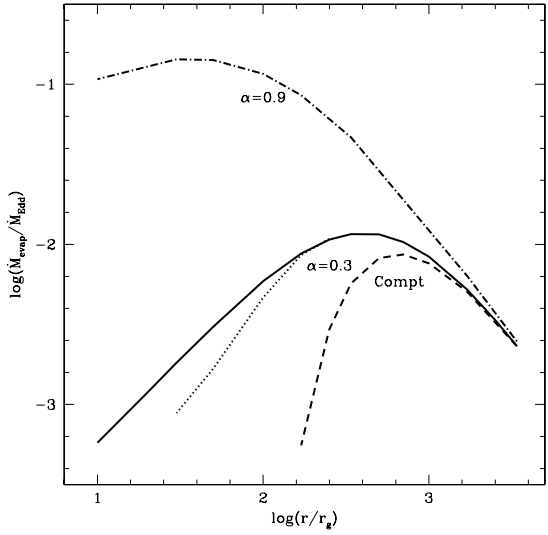}
\caption[]{\label{f:eta} Influence 
of mass inflow from the hump region on  the  evaporation.
Solid line: basic  case ($\alpha=0.3$ without Compton cooling). Dotted
line:  influence of hump flow.
Dashed line: influence of Comptonization.
Dash-dotted line: influence of $\alpha$.}
\end{figure}

Computations have shown that the coronal flow comes mainly from the dominant
evaporation region, i.e. the hump zone. As a rough estimate, we
simplify $\eta(r)$ to 
\begin{equation}\label{e:eta}
\eta(r)\approx {\dot M_{\rm evap}(r)\over \dot M_{\rm c}(r_{\rm hp})}.
\end{equation}
Replacing  $\dot M_{\rm evap}(r)$ in Eq.(\ref{e:eta}) by
$2\pi r^2 \dot m_0(r)$, we recalculate the coronal structure with the new radial
mass flux and energy flux,  the  evaporation
rate along distance is plotted in Fig.\ref{f:eta}. From the figure we see that 
the evaporation by including the inflow from outer neighboring region
is still inefficient in the inner region.
The characteristic feature of hump-evaporation efficiency
along distances doesn't change. 
Since $\eta$ is very small at small distances far from the hump region,
 the dependence of the evaporation rate on the radial advective loss is
quite weak. The coronal flow inside the hump is the same order of
magnitude as the hump flow. The evaporation rate and the coronal
flow seem to depend more strongly  on the detailed cooling and
heating processes, such as Comptonization, additional heating (large $\alpha$).
On this more consistent global investigations are needed but they
are beyond the scope  of this paper.

\subsection{Accretion via corona, disk, or both?} 
We investigated the hot corona lying above a cool accretion
disk and the mass and energy exchange between the corona and the
disk. Continuously mass  evaporates from the disk into the corona and flows
in the corona towards the central black hole. At large distance from
the black hole only accretion via a thin disk occurs. Further inward
(at distances of $10^3$ to $10^4$ Schwarzschild radii 
for X-ray transients in quiescence or low luminosity galactic nuclei) the accretion
can change to the mode of an ADAF (e.g. Narayan, McClintock, \& Yi 1996). 
This transition occurs where 
the evaporation rate exceeds the mass inflow rate in the cool disk.
 From that distance inward the thin disk is
depleted and the coronal flow fills the inward region up to the last
stable orbit. Consequently, the accretion changes from
thin disk outside to coronal flows/ADAFs inside. The transition radius
is determined by the values of mass
evaporation rate and mass flow rate in the disk.
Actually outside the transition radius both exist together, 
the cool disk and the corona above.
Such a truncation of the thin disk only occurs if the mass supply
rate in the disk is less than the maximal evaporation rate.
If the mass inflow rate is higher, the disk cannot be fully depleted
and we expect an interior disk below the corona inside the distance of
maximal evaporation. 
Summarizing these interactions, we claim that
the mass supply rate $\dot M_{\rm in}$, compared to $\dot M_{\rm
evap}$,
 determines
whether accretion via the corona or the disk, or both dominate: 
1) At low accretion rate $\dot M_{\rm in}$, a large part of the disk is depleted 
by mass evaporation 
and filled by coronal accretion flows/ADAFs, so accretion is dominated by
hot coronal flows/ADAFs.
 2) At a rate around the maximal evaporation rate the disk truncation 
reaches its innermost possible distance, i.e. the hump distance, and
an inner corona/ADAF coexists with
an outer disk.   
3) At a higher rate, the disk extends to the last stable orbit, and
the corona, which is negligible in the far outward parts, could be
also weak at innermost region due to the efficient Compton cooling.   
In such a way, the thin disk exists at all distances with a corona lying above. 

Up to now the evaporation model was applied to the situation of a thin
disk truncated at a certain distance, that is for mass flow rates 
$\dot M_{\rm in}$ less than the maximal $\dot M_{\rm evap}$.
(Meyer et al. 2000b, Liu \& Meyer-Hofmeister 2001).
Now the new computation of coronal structure for inner regions
and the evaporation efficiency there connect to the interesting 
questions how the radiation from these coronae with disks underneath
affects the spectrum.

We have shown that the evaporation rate strongly depends on 
the viscosity value,
especially that the maximum rate is changed and this maximum rate occurs 
for the higher viscosity at smaller distance than for lower $\alpha$.
The larger the viscosity value is, the smaller is the inner region
with an ADAF. 
Furthermore, the Compton cooling can affect the evaporation.
If the mass flow rate $\dot M_{\rm in}$ is high, the Compton cooling 
reduces the evaporation, and due to the reduced maximal 
evaporation rate the disk extends further inwards and may even extend  
to the last stable orbit.
Therefore, the dominant accretion mode and the transition radius 
from outer disk accretion to an inner ADAF sensitively depend on both,
viscosity and mass supply rate to the disk.   
The dependence of the transition radius on the mass accretion rate is
different from that in our previous investigation if Compton
cooling becomes important. 

\subsection{Corona + thin disk in AGN?}

In AGN hard X-ray tails and iron $K_\alpha$ lines are observed (e.g. Pounds
et al. 1990; Matsuoka et al. 1990; Tanaka et al. 1995),
implying that cold and hot gas
exist in the vicinity of the central black hole. Commonly it is 
believed that these components 
are corona and disk respectively. Soft photons from the disk are
Compton up-scattered by energetic electrons in the
optically thin corona and form the hard X-ray power-law spectral component.
The iron lines are from the illumination of disk matter by X-ray
radiations from the corona (e.g. Guilbert \& Rees 1988; Lightman \&
White 1988; George \& Fabian 1991).

The disk corona evaporation model shows that such kind of structure
can  exist self-consistently, if the mass accretion rate and the
viscosity have  appropriate values. 
Considering the high
luminosity in AGN, Comptonization is quite strong and the 
evaporation is restrained. 
Near the black hole part of the coronal gas is
expected to cool down and fall back to the disk.
To keep the corona  above the thin disk at this innermost region,
very large viscosity is required as is indicated in Fig.~\ref{f:mdot-r}.
This implies that a large
fraction of accretion energy needs to be released in the corona.
 Here we raise the same question as Merloni \& Fabian (2001b) did, that
how the corona stores and releases so much energy. 

In luminous AGN (i.e. large $\dot M/\dot M_{\rm Edd}$), corona can not exist in the innermost region anyhow as
Compton cooling is so strong that coronal plasma  completely cools
down and falls back to the disk. No strong hard X-ray spectral
component  is expected from corona.
This is consistent with the observational feature that 
high-$\dot{m}$ objects (e.g., Narrow-Line Seyfert
  1 galaxies  with
relatively small black hole mass)
 show steep hard X-ray spectra (e.g. Boller, Brandt, \& Fink 1996; 
Brandt, Mathur, \& Elvis 1997), 
while the relatively low luminosity AGN
(e.g., classical Broad-
  Line Seyfert 1s) show hard X-ray tails.

\section{Conclusions}

We studied the properties of disk and corona around black holes 
 including the exchange of mass and energy between the disk and
the corona. 
The decoupling of electron and ion temperatures and Compton cooling were taken
into account. We found that the viscosity parameter $\alpha$ for the coronal gas
has a strong influence on the evaporation efficiency, for larger $\alpha$
the mass flow in the corona increases and 
 the maximum of the evaporation rate is reached at a smaller
radius.
We also found that mass evaporation to the corona is largely 
restrained by Compton cooling at regions close to the black hole.
 Both
effects, larger viscosity and smaller incoming mass accretion rate to the disk,
lead to  the coronal flow/ADAFs extending outwards to a larger distance.
Coronal radiation is weak for high accretion rates, consistent with
the steep hard X-ray 
spectra observed in NLS1s and galactic black hole candidates. 
To maintain a corona lying above a thin disk down to the last stable
orbit in AGN, a
large energy store/release is required.   
The results based on galactic nuclei are mass independent and
therefore also applicable to
black hole X-ray binaries.

\acknowledgements{BFL and TK would like to thank the Japan Society for the
Promotion of Science for support. This work is partially supported by
the Grants-in Aid of the Ministry of Education, Science, Sports, and
Culture of Japan (P.01020, BFL; 13640238, SM; P.4616, TK).}

\onecolumn

\twocolumn


\end{document}